\documentclass[a4paper]{jpconf}
\usepackage{graphicx}
\begin{document}
\begin{flushright}
UCLA/05/TEP/32\\
December 2005
\end{flushright}

\vspace{-1.5truecm}

\title{GZK photons  as UHECR above 10$^{19}$~eV\footnote{Talk given at TAUP2005, Sept. 10-14 2005, Zaragoza (Spain)}}

\author{Graciela B Gelmini}

\address{Department of Physics and Astronomy, UCLA, Los Angeles,
CA 90095-1547, USA}

\ead{gelmini@physics.ucla.edu}

\begin{abstract}
``GZK photons" are produced by extragalactic nucleons through the
resonant photoproduction of pions. We present the expected range
of the GZK photon fraction of UHECR, assuming a particular UHECR
spectrum and primary nucleons, and compare it with the
minimal  photon fraction predicted by Top-Down models.
\end{abstract}
The Pierre Auger Observatory~\cite{Auger} may prove photon fractions in the Ultra-High Energy Cosmic Rays
(UHECR) above $10^{19}$~eV at the level of 10\% (maybe even a few\%).
Based on Ref.~\cite{GKS}, here  we will  address the physical implications of such detection
or limit. In particular, we will discuss if ``GZK photons" could be observed at this level or, otherwise, if limits
on the parameters on which their flux depends could be obtained from the non observation
of photons at this level. We call ``GZK photons" those photons produced by  extragalactic nucleons through the resonant photoproduction of pions, the so called GZK effect.
In Ref.~\cite{GKS} we fitted the assumed UHECR spectrum above  $2 \times10^{19}$~eV
 solely with primary nucleons and the GZK photons they produce. The GZK photon flux 
  depends on the UHECR spectrum assumed,  the slope and maximum energy of
the primary nucleon spectrum, the minimum distance to the sources and the intervening radio background
and average extragalactic magnetic field.
We took a phenomenological approach in choosing the range of the several relevant
parameters, namely we took for each of them a range of values mentioned in the literature, 
without attempting to assign them to particular sources or acceleration mechanisms.
We also assumed the existence of a galactic or extragalactic   Low Energy Component (LEC) when necessary to fit the
assumed UHECR at energies below $10^{19}$~eV, taking care that it is
negligible at energies  3$\times 10^{19}$~eV and above.

We used a numerical code developed in Ref.~\cite{kks1999}  
to compute the flux of GZK photons produced by an homogeneous distribution of sources emitting originally
only protons (however the results at the high energies considered is the same for primary neutrons).  It  uses the  kinematic equation approach and
calculates the propagation of  nucleons, stable leptons  and photons
 using the standard dominant processes.

 We parametrized the initial proton flux for any source with a power
law function, $F(E) = f~ E^{-\alpha}~ \theta(E_{\rm max} -E)$.
The power law index $\alpha$ and maximum energy $E_{\rm max}$ were considered free
parameters. The amplitude $f$ was fixed by normalizing the final proton
flux from all sources to the observed flux of UHECR, which we took to be 
either the AGASA spectrum or the HiRes spectrum
(given that a reliable Auger spectrum in not yet
available). We considered the
 power law index to be in the  range  $1 \le \alpha \le 2.7$ and
 $E_{\rm max}$ between $10^{20}$~eV and $10^{22}$~eV.
 For the average extragalactic magnetic field we took the range $10^{-11}$G to
   $10^{-9}$G,~\cite{egmf}
 and for the radio background we considered three estimates: one from Clark et al. and two higher ones
  from Protheroe and Biermann~\cite{radio}.

The largest GZK photon fractions in UHECR
 happen for small values of $\alpha$, large values of $E_{\rm max}$,
  small minimal distance to the sources (which is compatible with a
small frequency of clustering of the events) and small intervening backgrounds. The smallest
GZK photon fluxes are obtained with the opposite choices.
In the most favorable
cases for a large photon flux, GZK photons could dominate the UHECR flux in an
energy range above 
10$^{20}$~eV. This allowed us~\cite{GKS}
to fit the AGASA data, at the expense of
assuming that the initial protons could have  a hard spectrum $\sim 1/E$
and be accelerated to energies as high
 as $10^{22}$~eV. In this extreme case, the AGASA data can be explained
 without any new physics, except in what the mechanism of acceleration of
 the initial protons  is concerned. With the HiRes spectrum the GZK photons are always subdominant
and  can be neglected for the fit.

\begin{figure}[ht]
\includegraphics[width=0.48\textwidth,clip=true,angle=0]{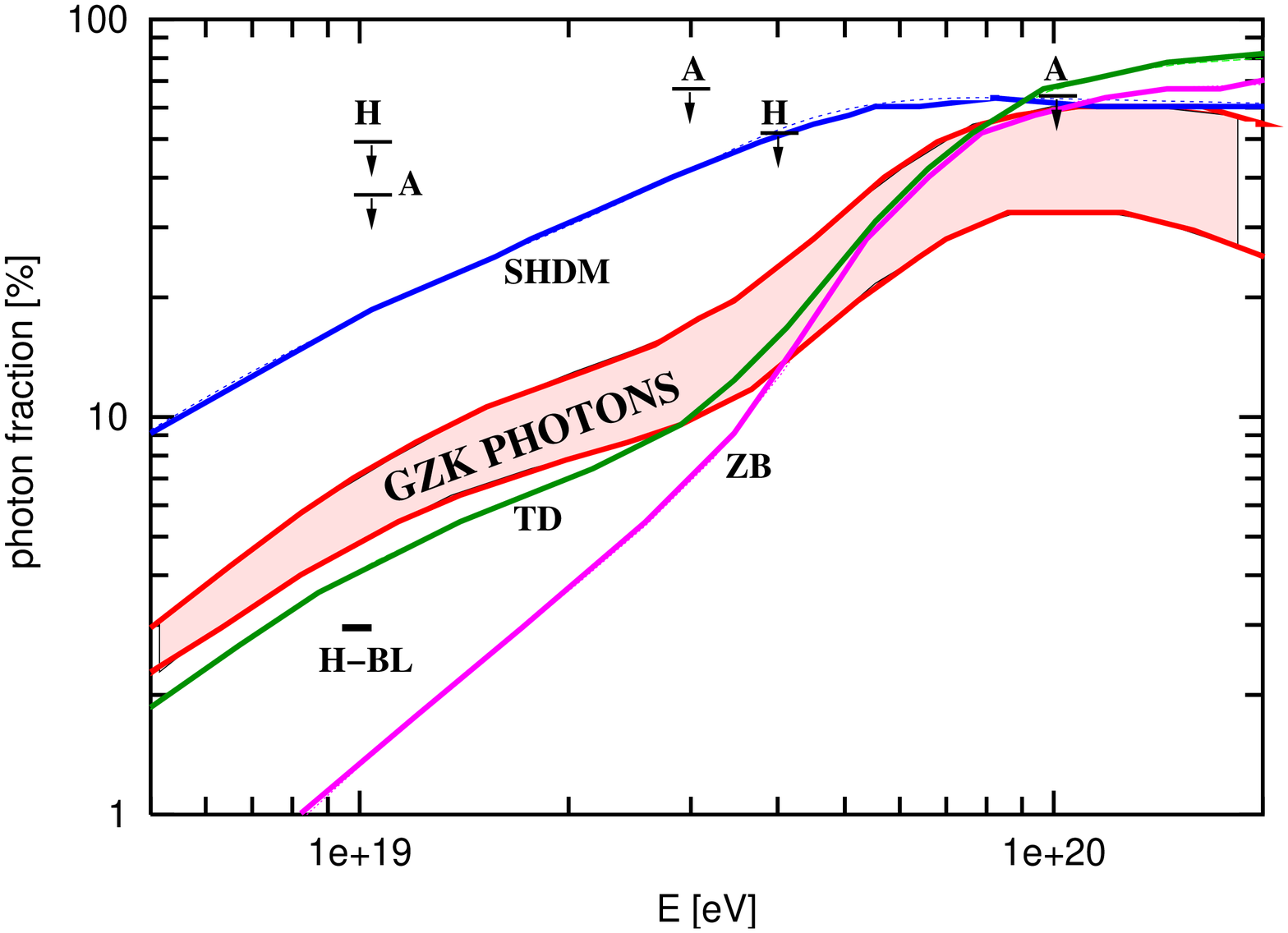}
\includegraphics[width=0.5\textwidth,clip=true,angle=0]{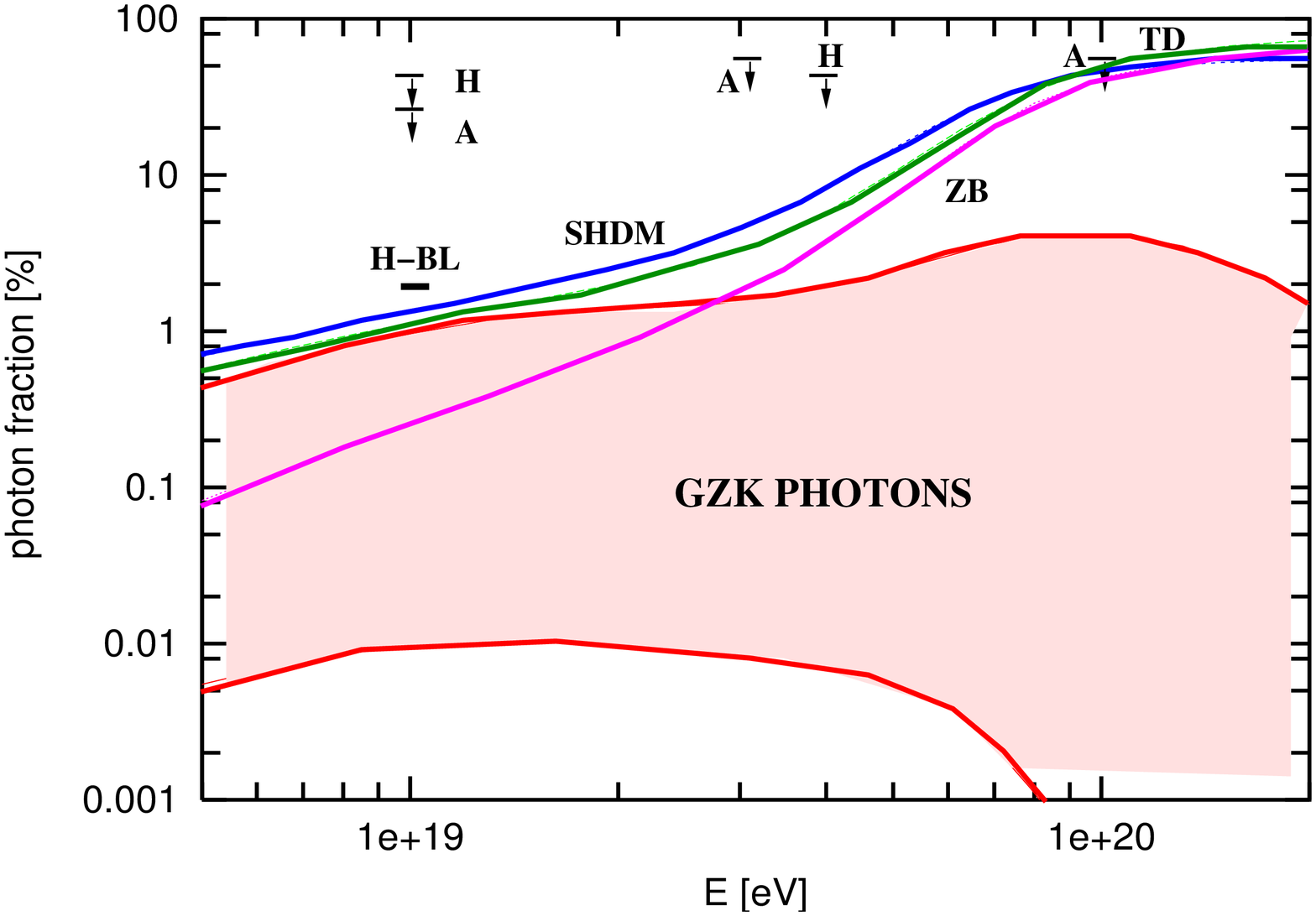}
\caption[...]{Photon fraction in percentage of the total 
 predicted integrated UHECR spectrum
above  the energy $E$ for (a) the AGASA spectrum (left panel) and (b)
the HiRes spectrum (right panel). The pink regions show the range of GZK photon
fractions expected if only nucleons are produced at 
the sources. The curves labeled
 ZB (Z-bursts), TD (topological defects- necklaces) and
SHDM  (Super Heavy Dark Matter model) show examples of
 minimum photon fractions predicted by these models.
Upper limits: {\bf A} from  AGASA, Ref.~\cite{agasa_composition_2} 
at $1-3\times 10^{19}$~eV, and,
Ref~\cite{agasa_photon}, obtained with AGASA data at $10^{20}$ eV;
 {\bf H} from Haverah Park~\cite{haverah};   {\bf H-BL}
show the fraction of HiRes stereo events required to explain a correlation
with BL Lac sources~\cite{hires_neutral}.}
\label{F1}
\end{figure}

 Proceeding in this manner, in Ref.~\cite{GKS}
we fitted  the AGASA~\cite{agasa} and  the HiRes monocular~\cite{hires} data 
 trying to minimize and to maximize the number of GZK protons produced, to obtain the expected range
 of the GZK-photon fraction in UHECR.
We found (see  the pink bands  in Fig.~\ref{F1}) that the GZK photon fraction of the total integrated  UHECR flux, for the AGASA spectrum
is between 5\% and 7\%  above $10^{19}$ eV  and between 30\% and 60\% above
$10^{20}$ eV, thus Auger should be able to see these photons or place interesting bounds on the flux parameters. Recall that fitting the AGASA data with astrophysical sources requires the extreme choices for the initial spectrum mentioned above.  With the HiRes spectrum, instead
 the predicted GZK photon fraction is between  0.01\%  and 1\% of
 the UHECR  above $10^{19}$ eV  and between  0.001\%  and  4\% 
  above $10^{20}$ eV, thus these photons may or may not be within the reach of Auger.

Detection of these GZK photons would open the way for  UHECR photon astronomy.
Detection of a larger photon flux than expected for GZK photons given the
particular UHECR spectrum assumed,
would imply the emission of photons at the
source or new physics. New physics is involved in Top-Down models,
proposed as an alternative to acceleration models to explain the
 origin of the highest energy cosmic rays. All  Top-Down models
 predict photon dominance at the highest energies (and no heavy nuclei).
In Ref.~\cite{GKS}, we estimated  the minimum photon
fraction Top-Down models predict, not only assuming the AGASA spectrum,
which these models were originally proposed to explain,  but
also assuming the HiRes spectrum.  We  showed that at high energy, close to
10$^{20}$~eV,
the maximum expected  flux of GZK photons is comparable to (for the
 AGASA spectrum)
or much smaller than (for the HiRes spectrum)  the minimum flux of
 photons predicted
by Top-Down models which fit the AGASA or the HiRes data (see Fig.~\ref{F1}).
 Three Top-Down models were considered:  Z-bursts~\cite{zburst},
 topological defects (necklaces) (for a review see for example Ref.~\cite{td_review}) and super
heavy dark matter particles~\cite{SHDM}
(in particular,  predictions of
 Ref.~\cite{SHDM_2004} were used).

 In order to estimate the minimum   photon
ratio  predicted by Top-Down models we assumed that these models explain only
the highest energy UHECR (if they do not explain even those events,
 the models are irrelevant
for UHECR). For this purpose we made liberal use of a component of extragalactic nucleons
which would explain all but the highest energy UHECR. The
assumption of two components, one of accelerated nucleons and another of
Top-Down generated particles, which would conspire to produce a continuous spectrum, is contrived, and was used  only as a means to minimize
the number of Top-Down photons. Fitting the AGASA and HiRes  UHECR spectra in this manner we showed that the photon ratio
at energies close to $10^{20}$ eV  is always larger than  10\%,
 in most cases is larger than 50\%,
 independently of the UHECR spectrum assumed, making 
this is a crucial test for Top-Down models. Reaching the level of $\sim$30\% (10\%) on the photon ratio
at energies close to $10^{20}$ eV, Auger will find UHECR photons or reject most (all)  Top-Down models, independently of 
the UHECR spectrum.

\section*{Acknowledgments}

This work  was supported in part by NASA grant NAG5-13399.
G.G was supported in part by the US DOE grant DE-FG03-91ER40662 
Task C.

\end{document}